\documentclass[prb,twocolumn,amsmath,amssymb]{revtex4-1}

\usepackage{graphicx}
\usepackage{dcolumn}
\usepackage{bm}
\usepackage[colorlinks=true]{hyperref}
\hypersetup{linkcolor=blue}
\usepackage{algorithm} 
\usepackage{algorithmic}
\usepackage{verbatim}

\begin{document}

\title{Mott insulators of ultracold fermionic alkaline earth atoms in three dimensions}

\author{Hao Song}

\author{Michael Hermele}

\affiliation{Department of Physics, University of Colorado, Boulder, Colorado
80309, USA}
\begin{abstract}
We study a class of $SU(N)$ Heisenberg models, describing Mott insulators of fermionic ultra-cold alkaline earth atoms on the three-dimensional simple cubic lattice.  Based on an earlier semiclassical analysis, magnetic order is unlikely, and we focus instead on a solvable large-$N$ limit designed to address the competition among non-magnetic ground states. We find a rich phase diagram as a function of the filling parameter $k$, composed of a variety of ground states spontaneously breaking lattice symmetries, and in some cases also time reversal symmetry.  One particularly striking example is a state spontaneously breaking lattice rotation symmetry, where the cubic lattice breaks up into bilayers, each of which forms a two-dimensional chiral spin liquid state.
\end{abstract}
\maketitle

\section{Introduction}

Ultracold atom experiment techniques enable us to vary parameters 
of quantum many-body systems that can hardly
be changed in solid state materials.\cite{jaksch2005, lewenstein2007, bloch2008}
For example, in solid
state systems the crystal structure is selected by nature, so it is
usually not easy to study the dependence of the system properties
on the lattice structure. But in ultracold atom experiments the optical
lattice can be chosen artificially, and its dimension and geometry
can be varied. Also, we
have significant freedom to select the constituent particles of a
many-body system. They can be atoms or molecules, bosons or fermions, and so on.
Different atoms or molecules interact
with one another quite differently, and in some cases the interactions
can be tuned with electric or magnetic field. So cold atoms promise
to allow us explore systems in new parameter regimes, or even systems
that have no analog in solid state materials.

Fermionic\cite{bosonicnote} 
ultracold alkaline earth atoms (AEAs) have attracted significant interest
recently due to their unique properties,\cite{gorshkov2010twoorbital,cazalilla2009,hermele2009mottinsulators,fossfeig2010a,fossfeig2010b,xuc2010,toth2010,corboz2011simultaneous,hermele2011topological,manmana2011,rapp2011,szirmai2011gaugefields,bauer2012,bonnes2012,messio2012,tokuno2012,hazzard2012hightemperature,cai2012pomeranchuk,cai2012quantummagnetic,corboz2012simplex,corboz2012spinorbital} and experimental
progress developing the study of many-body physics in AEA systems has been rapid \cite{fukuhara2007a, fukuhara2007b, fukuhara2009, escobar2009, stellmer2009, desalvo2010, taie2010, tey2010, stellmer2010, sugawa2011,blatt2011,bishof2011a,bishof2011b, lemke2011,stellmer2011,stellmer2012}.
One key feature of AEAs is
the presence, to an excellent approximation, of $SU(N)$ spin rotation
symmetry, where $N=2I+1$ and $I$ is the nuclear spin.\cite{gorshkov2010twoorbital, cazalilla2009}
 This occurs
in both the $^{1}S_{0}$ ground state and a metastable $^{3}P_{0}$
excited state, where the electronic angular momentum $J_{e}=0$ and
the hyperfine interaction is thus quenched. This leads to the nuclear-spin-independence
of the s-wave scattering lengths between AEAs, and to $SU(N)$ spin
rotation symmetry. When loaded in optical lattices, AEA systems are
 described by $SU(N)$-symmetric Hubbard models.\cite{gorshkov2010twoorbital} 
Since the largest $I$
obtained using AEA is $I=9/2$ in the case of $^{87}Sr$, $N\leq10$
is the experimentally accessible regime. Different setups are possible, and
as a result,  $SU(N)$ versions of several models, such as the Kugel-Khomskii
model, the Kondo lattice model, and the Heisenberg spin model, can be realized
with AEAs as special or limiting situations of the more general Hubbard model.

Among these models, we focus in this paper on $SU(N)$ antiferromagnetic
Heisenberg models, which describe the Mott insulator phase of fermionic
AEAs in optical lattices. More specifically, we are concerned with such models
on three dimensional lattices, which have received much less attention
than the one- and two-dimensional cases.  Because of the enlarged symmetry, the number
of spins needed to make a singlet, denoted by $k$, is in general
larger than two. In the simplest AEA Heisenberg model with one atom per lattice site, $k = N$.
In addition, in the semiclassical limit of the Heisenberg
models that can be realized using AEAs, two neighboring classical spins
prefer energetically to be \emph{orthogonal} rather than anti-parallel.\cite{hermele2009mottinsulators}
Both these features contrast with $SU(2)$ antiferromagnetic Heisenberg models
appropriate for some solid state materials, where neighboring \emph{pairs} of spins can
and tend to form singlet valence bonds, and neighboring classical spins prefer to be anti-parallel.
We can thus expect new physics in $SU(N)$ Heisenberg models with $k > 2$.  

Indeed, Ref.~\onlinecite{hermele2009mottinsulators} argued that the underconstrained nature
of the semiclassical limit makes magnetic order unlikely for large enough $N$ on any lattice,
and non-magnetic ground states are more likely.  While the models of physical interest are challenging
to study directly, information about possible non-magnetic ground states can be obtained in a large-$N$
limit designed to address the competition among such states.\cite{affleck1988largenlimit,marston1989largenlimit,read1989somefeatures}
Such a large-$N$ study was carried out for AEA $SU(N)$ Heisenberg models on the two-dimensional square lattice in
Refs.~\onlinecite{hermele2009mottinsulators,hermele2011topological}.   One possible
non-magnetic state is a cluster state, where clusters of $k$ (or a multiple
of $k$) neighboring spins form singlets; this is a generalization of
a valence bond state. Another possibility is a spin liquid state,
where full translational symmetry is preserved.  For the simplest AEA Mott insulators
(with $^{1}S_0$ ground state atoms only), on the square lattice the large-$N$ study finds
cluster states for $k \leq 4$, and a chiral spin liquid (CSL) state for $k \geq 5$.\cite{hermele2009mottinsulators,hermele2011topological}
The CSL spontaneously breaks time-reversal ($\mathcal{T}$)
and parity ($\mathcal{P}$) symmetries, and can be viewed as a magnetic analog
of the fractional quantum Hall effect (FQHE), with similar exciting properties of
quasiparticles with anyonic statistics, gapless chiral edge states, and so on.\cite{kalmeyer1987, kalmeyer1989, wen1989} CSLs have  also been found in a variety of other exactly solvable models.\cite{khveshchenko1989, khveshchenko1990, yao2007, schroeter2007, greiter2009, thomale2009, greiter2011}

The CSL is, however, intrinsically a two-dimensional
phenomenon, so it is natural to ask about non-magnetic ground
states of  $SU(N)$ antiferromagnetic Heisenberg models in three dimensions.  In this paper, we address this question by a large-$N$ study of a class of $SU(N)$ Heisenberg models
on the simple cubic lattice, and find a rich phase diagram as a function of $k$ including cluster states,
but also more intricate inhomogenous states.  Most strikingly, for $k=7, 10$ we find a bilayer CSL state, where the lattice
spontaneously breaks into weakly coupled square bilayers (thus breaking rotational symmetry), 
each of which is a two-dimensional CSL.  We thus find that the CSL survives to three dimensions, relying on spontaneous symmetry breaking that results in effective quasi-two-dimensionality.

We now define our model before briefly surveying some related prior work.  We consider a fermionic AEA with $N$ spin species, and put $m$ $^{1}S_0$ ground state atoms on each site of a simple cubic lattice (see Sec.~\ref{sec:model} for more details).  The atoms form a Mott insulator due to repulsive on-site interactions.  For simplicity, we consider  the case of dominant on-site interaction, so that the spin degrees are governed by a antiferromagnetic superexchange interaction restricted to nearest neighbors.  While $m=1$ is the most interesting
situation since it best avoids three-body losses, we also consider more generally the case where 
$\frac{N}{m}$ is an integer. Then, the minimum number
of spins needed to make a $SU(N)$ singlet is $k=\frac{N}{m}$.  We sometimes refer to $k$ as the filling parameter.  When $m=1$, each spin
transforms in the fundamental representation of $SU(N)$.  In the large-$N$ limit, $N$ is taken large
while $k$ is held fixed.  Given the physical interpretation of $k$, we thus view the large-$N$ results for a given $k$ as a guide to the physics of the physically realizable model with $m = 1$ and $N = k$.

Our focus is on three spatial dimensions, but we note that one-dimensional $SU(N)$ Heisenberg spin chains have been solved exactly for the case $m=1$,\cite{sutherland1975} and the effective field theory of such chains is understood for general $m$.\cite{affleck1988criticalbehavior}  The latter analysis shows that gapless states with quasi-long-range order, as well as gapless cluster states, occur in one dimension.  In two dimensions, early studies of $SU(N)$ antiferromagnets focused on models where two neighboring spins can be combined to form a singlet.  This work included the models we consider for the case $m = N/2$,\cite{affleck1988largenlimit,marston1989largenlimit}
but also other $SU(N)$ antiferromagnets with spins transforming in two distinct conjugate  representations on the two sublattices of a bipartite lattice.\cite{read1989somefeatures}  Models with $k=2$ have also received attention more recently,\cite{honerkamp2003,assaad2005,xuc2010,cai2012quantummagnetic} and two dimensional models with $k > 2$ have been studied\cite{pokrovskii1972,li1998,bossche2000plaquette,lauchli2006,arovas2008simplex,wangf2009,hermele2009mottinsulators,toth2010,corboz2011simultaneous,rapp2011,hermele2011topological,corboz2012simplex,corboz2012spinorbital} (see Ref.~\onlinecite{hermele2011topological} for a more detailed discussion of some of these prior works).  The $m=1$, $N=3$ model on the square lattice is magnetically ordered,\cite{toth2010} and there is also evidence for magnetic order for $m=1$, $N=4$.\cite{corboz2011simultaneous}  Only a little attention has been devoted to the case of three dimensions,\cite{pokrovskii1972,fukushima2005,toth2010} but we note the high temperature series study of Ref.~\onlinecite{fukushima2005}, where the $m=1$ model on the simple cubic lattice was studied for various values of $N$, and it was found that increasing $N$ led to a decreased tendency toward magnetic order.   References~\onlinecite{pankov2007,xuc2008} studied effective models for four-site singlet clusters on the cubic lattice.  Finally, we  note that high-spin quantum magnets can also be realized using ultra-cold alkali atoms.  While $N$-component such systems do not generically obey $SU(N)$ spin symmetry, the symmetry is enhanced above $SU(2)$,\cite{wucj2003} and such systems have received significant attention.\cite{wucj2003,chen2005,lecheminant2005, wucj2005, wucj2006, wucj2010,szirmaie2011} 

In Sec.~\ref{sec:model}, we review the large-$N$ solution to our model.  This is followed by presentation of the large-$N$ results for $k = 2,\dots,10$ in Sec.~\ref{sec:largen}, together with a discussion of how those results are obtained and checked.  As part of that discussion, we develop an interesting relation between some cubic lattice saddle points (including the ground state saddle points for $k = 5,\dots,10$) and saddle points on the single-layer square lattice with filling parameter $k' = k/2$.  The paper concludes with a discussion of the striking properties of the bilayer CSL state (Sec.~\ref{sec:discussion}).

\section{Theoretical Model}
\label{sec:model}

The $SU(N)$ Hubbard model 
\begin{eqnarray}
{\cal H}_{Hubbard} & = & -t\sum_{\left\langle r r'\right\rangle }\left(c_{r}^{\alpha\dagger}c_{r'\alpha}+h.c.\right)\nonumber \\
\lefteqn{} &  & +\left(U/2\right)\sum_{r}\left(c_{r}^{\alpha\dagger}c_{r\alpha}-m\right)^{2},\label{eq:Hubbard}
\end{eqnarray}
describes the behavior of fermionic AEAs on an optical lattice.\cite{gorshkov2010twoorbital} Here
$c_{r}^{\alpha\dagger}$ and $c_{r\alpha}$ are the creation and annihilation
operators for the fermionic atom with spin state $\alpha$ at site
$r$. The sum in the first term is over nearest-neighbor pairs of lattice sites.  We will primarily consider 
the simple cubic lattice.
We choose the number of atoms so that $m$ is the integer number
of atoms per lattice site. There are $N$ spin states, $\alpha,\beta=1,2,\dots,N$,
and spin indices are summed over when repeated.  The total number of lattice sites is $N_s$. The operator $c_{r}^{\alpha\dagger}$ transforms in the fundamental representation of $SU(N)$, while $c_{r\alpha}$ transforms in the anti-fundamental representation, which is related to the fundamental by complex conjugation.  The upper and lower
positions of the Greek indices are used to indicate the distinction between these two representations (they are unitarily equivalent only for $N=2$).

As is well known, the $SU(2)$ Heisenberg model can be obtained
as a low energy effective description of the $SU(2)$ Hubbard model
when $U\gg t$. The generalization to the $SU(N)$ version is straightforward.
In second order degenerate perturbation theory, one obtains the $SU(N)$ antiferromagnetic
Heisenberg model defined by the Hamiltonian 
\begin{equation}
\mathcal{H}=-J\sum_{\left\langle r r'\right\rangle }(f_{r}^{\alpha\dagger}f_{r'\alpha})(f_{r'}^{\beta\dagger}f_{r\beta}),\label{eq:Hamiltonian-1}
\end{equation}
 with the Hilbert space restricted by $f_{r}^{\alpha\dagger}f_{r\alpha}=m$, and $J = 2 t^2 / U > 0$.  We now use $f_{r}^{\alpha\dagger}$ rather than $c_{r}^{\alpha\dagger}$ to denote the fermion creation operator, to emphasize that once we pass to the Heisenberg model, the fermions do not move from site to site.  This is important, because the structure of the large-$N$ mean-field theory is that of a hopping Hamiltonian for the $f_{r}^{\alpha\dagger}$ fermions, but it is not correct to interpret this hopping as motion of atoms.  Instead, in the large-$N$ mean-field theory, the $f_{r}^{\alpha\dagger}$ fermions are spinons, fractional particles that may be either confined or deconfined depending on the nature of fluctuations about a mean-field saddle point.  See Ref.~\onlinecite{hermele2011topological} for further discussion of this point.
 
On each site, there are $m$ atoms that form a $SU(N)$
spin. The Hamiltonian~(\ref{eq:Hamiltonian-1}) defines an antiferromagnetic
interaction, since by rearranging the fermion operators it can be
written as 
\begin{equation}
{\cal H}=J\sum_{\left\langle r r'\right\rangle }\hat{S}_{\alpha}^{\beta}(r)\hat{S}_{\beta}^{\alpha}(r'),\label{eq:SpinH}
\end{equation}
 where $\hat{S}_{\alpha}^{\beta}(r)=f_{r}^{\beta\dagger}f_{r\alpha}$
flips the spin on site $r$.

We study this model on the simple cubic lattice, the simplest three dimensional
case, with varying parameters $N$ and $m$.
While we consider more
general parameter values, $m=1$ is the case of greatest physical interest
because putting
only one atom on each site best avoids potential issues due to three
body loss. The largest $N$ that can be obtained using alkaline earth atoms
is $N = 10$ in the case of $^{87}Sr$.  

Based on a semiclassical analysis, Ref.~\onlinecite{hermele2009mottinsulators} argued that for large enough $N$, 
magnetic ordering is unlikely on any lattice.  The argument proceeds in the semiclassical limit, where a lower bound on the dimension of the ground state manifold is derived.  For $N > N_c$, where $N_c$ depends on the lattice coordination number, the ground state manifold is extensive, meaning its dimension is proportional to the number of lattice sites.  This situation occurs in some geometrically frustrated systems and is likely to lead to a strong or complete suppression of magnetic order\cite{moessner1998}, even in the semiclassical limit that favors magnetic order by construction.  Therefore, non-magnetic ground states are likely when $N > N_c$.  For the square lattice $N_c = 3$,\cite{hermele2009mottinsulators} and the argument is easily extended to find $N_c = 4$ on the cubic lattice.

Ideally, we would like to predict the properties of the $SU(N)$ antiferromagnetic
Heisenberg model on cubic lattice for $N\leq10$, $m=1$. But this
is extremely challenging.  Instead, following the work of Refs.~\onlinecite{hermele2009mottinsulators,hermele2011topological},
we apply a large-$N$ limit in which the model becomes exactly
solvable, and which allows us to address the competition among different
non-magetic ground states. We fix the ratio $k=\frac{N}{m}$ (for
integer $k$), while taking both $N\to\infty$ and $m\to\infty$. We shall sometimes refer
to $k$ as the filling parameter.  For
each $k$ we thus obtain a sequence of models $(N=k, m=1)$; $(N=2k,
m=2)$, and so on. For every model in this sequence, $k$ is the minimum
number of spins needed to form a singlet, and it is thus reasonable
 that the large-$N$ limit may capture the physics of the
case $N=k$, $m=1$ of greatest interest. 

To proceed with the large-$N$ solution, one goes to a functional
integral representation, where the partition function is 
\begin{equation}
Z=\int{\cal D}\chi{\cal D}\chi^* {\cal D}\lambda{\cal D}\bar{f}{\cal D}f\, e^{-S},\label{eq:PartionF}
\end{equation}
 where 
\begin{eqnarray}
S & = & \int_{\tau}\sum_{r}\bar{f}_{r}^{\alpha}\partial_{\tau}f_{r\alpha}+N\int_{\tau}\sum_{\left\langle r r'\right\rangle }\frac{\left|\chi_{rr'}\right|^{2}}{{\cal J}}\nonumber \\
 &  & +\int_{\tau}\sum_{\left\langle r,r'\right\rangle }\left(\chi_{rr'}\bar{f}_{r}^{\alpha}f_{r'\alpha}+h.c.\right)\nonumber \\
 &  & +i\int_{\tau}\sum_{r}\lambda_{r}\left(\bar{f}_{r}^{\alpha}f_{r\alpha}-m\right).\label{eq:Action}
\end{eqnarray}
The field $\chi_{rr'}$ is a complex Hubbard-Stratonovich field that
has been used to decouple the exchange interaction, and $\lambda_{r}$
is a real Lagrange-multiplier field enforcing the $f_{r}^{\alpha\dagger} f_{r \alpha} =m$ constraint.
The fermion fields $f$ and $\bar{f}$ are the usual Grassmann variables.  We have introduced ${\cal J} = N J$; ${\cal J}$ is held fixed in the large-$N$ limit. 
Finally, $\int_{\tau}\equiv\int_{0}^{\beta}d\tau$. We shall always
be interested in zero temperature, \emph{i.e.} $\beta \to \infty$. 

When both $N$ and $m$ are large, the effective action for $\chi$
and $\lambda$ (obtained upon integrating out fermions), is proportional
to $N$ (since $m\thicksim N$), and therefore the saddle point approximation
becomes exact for the $\chi$ and $\lambda$ integrals. We can therefore
replace $\chi$ and $\lambda$ by their saddle-point values, $\chi_{rr'}\rightarrow\bar{\chi}_{rr'}$
and $\lambda_{r}\rightarrow i\mu_{r}$. The saddle-point equations are
\begin{eqnarray}
m & = & \left\langle f_{r}^{\alpha\dagger}f_{r\alpha}\right\rangle ,\label{eq:Saddle_1}\\
\bar{\chi}_{r r'} & = & -\frac{\cal{J}}{N}\left\langle f_{r'}^{\alpha\dagger}f_{r\alpha}\right\rangle .\label{eq:Saddle_2}
\end{eqnarray}
 The above averages are taken in the ground state of the saddle-point
(or mean-field) Hamiltonian 
\begin{eqnarray}
{\cal H}_{MFT} & = & N\sum_{\left\langle r r'\right\rangle }\frac{\left|\bar{\chi}_{rr'}\right|^{2}}{{\cal J}}+m\sum_{r}\mu_r \nonumber \\
 &  & +\sum_{\left\langle r r'\right\rangle }\left(\bar{\chi}_{rr'}f_{r}^{\alpha\dagger}f_{r'\alpha}+h.c.\right)-\sum_{r}\mu_{r}\hat{n}_{r},\label{eq:MFT}
\end{eqnarray}
 where $\hat{n}_{r}\equiv f_{r}^{\alpha\dagger}f_{r\alpha}$.

The ground state is determined by finding the global minimum of $E_{MFT}\left(\left\{ \chi_{rr'}\right\} ,\left\{ \mu_{r}\right\} \right)$,
the ground state energy of ${\cal H}_{MFT}$, as a function of the $\chi$'s
and $\mu$'s, with the constraint that the saddle point equations must be satisfied.   While any solution of the saddle point equations gives an extremum of the energy, in general it is not trivial to find the global minimum.  To address this question, we follow Refs.~\onlinecite{hermele2009mottinsulators,hermele2011topological} and apply the combination of analytical and numerical techniques developed there, as described below in Sec.~\ref{sec:largen}.

\section{Large-$N$ Ground States}
\label{sec:largen}

\subsection{Summary of the large-N mean-field results}
In the limit $N\to\infty$, the ground states are characterized entirely
by the mean-field saddle point values of $\chi_{r r'}$ and
$\mu_{r}$. The most important information is contained in
$\chi_{r r'}$, since typically it is possible for a given
$\chi_{r r'}$ to find $\mu_{r}$ so that the density
constraint Eq.~(\ref{eq:Saddle_1}) is satisfied. For instance, depending
on whether two sites are connected (i.e. whether there is a set of
nonzero $\chi_{r r'}$'s forming a path connecting the two
sites), we can tell whether the spins on the two sites are correlated
or not. Not all the information contained in $\chi_{r r'}$
is physical. The theory has a $U(1)$ gauge redundancy
\begin{equation}
\begin{array}{c}
f_{r\alpha}\rightarrow f_{r \alpha}e^{i\phi(r)}\\
\chi_{rr'}\rightarrow\chi_{rr'}e^{i\left(\phi(r)-\phi(r')\right)}
\end{array},\label{eq:GaugeTran}
\end{equation}
so the physical information is contained in the following gauge-invariant
quantities: (1) magnitude $\left|\chi_{rr'}\right|$ and (2) flux $\Phi=a_{12}+a_{23}+a_{34}+a_{41}$
through each plaquette, where 1, 2, 3, 4 indicates the four vertices
of a plaquette and $a_{rr'}$ is the phase of the $\chi_{rr'}$, \emph{i.e.}
$\chi_{r'r}=e^{ia_{rr'}}\left|\chi_{rr'}\right|$. (Since $\chi_{r'r}=\chi_{rr'}^{*}$,
$a_{r'r}=-a_{rr'}$.)

Based on a combination of analytical and numerical techniques described
below, we found the ground state configuration of $\chi_{r r'}$
and $\mu_r$ for $k=2,\dots,10$. These results, which
are rigorous for $k=2,3,4$, are summarized in Table~\ref{table:summary}. 
Different types of ground states are found
 depending on $k$. In an $n$-site cluster pattern of $\chi_{rr'}$,
the lattice is partioned into $n$-site clusters such that $\chi_{rr'}\neq0$
only if $r,r'$ lie in the same cluster. We call
the corresponding ground state a $n$-site cluster state, which can
be viewed as a generalization of a valence bond state (2-site cluster
state, in our terminology). Similarly, a bilayer pattern partitions
the lattice into bilayers, and $\chi_{rr'}$ is
only nonzero for $r,r'$ in the same bilayer. The
corresponding ground states are called bilayer states. In all cases,
each bilayer is comprised of two adjacent $\{100\}$ lattice planes.
A CSL bilayer is a special kind of bilayer state, where in each bilayer
\begin{equation}
\left|\chi_{rr'}\right|=\begin{cases}
\chi, & \left\langle rr' \right\rangle \text{ lies within either layer;}\\
\frac{{\cal J}}{k}, & \left\langle rr' \right\rangle \text{ connects the two layers.}
\end{cases}\label{eq:bilayer_1}
\end{equation}
Moreover, there is a uniform flux 
\begin{equation}
\Phi=\frac{4\pi}{k}\label{eq:bilayer_2}
\end{equation}
through each plaquette lying within the two layers, and zero flux
through each plaquette perpendicular to the two layers. This situation
corresponds to a uniform orbital magnetic field applied perpendicular to the
layers. At the mean-field level, a single CSL bilayer exhibits integer
quantum Hall effect with $\nu=1$ for each spin species of $f_{r\alpha}$
fermion.

\begin{table}
\begin{tabular}{|c|c|c|c|}
\hline 
$k$  & Large-$N$ ground state  & Sketch of $\chi_{r r'}$ & Energy\tabularnewline
\hline 
\hline 
2  & 2/4-site cluster  & Fig.~\ref{fig:cluster_k234}a & -0.125\tabularnewline
\hline 
3  & 6-site cluster  & Fig.~\ref{fig:cluster_k234}b & -0.0833333\tabularnewline
\hline 
4  & 4/8-site cluster  & Fig.~\ref{fig:cluster_k234}c & -0.0625\tabularnewline
\hline 
5  & 20-site cluster  & Fig.~\ref{fig:k5-10}a, \ref{fig:k5-10}b & -0.0445021 \tabularnewline
\hline 
6  & 12-site cluster  & Fig.~\ref{fig:k5-10}c, \ref{fig:k5-10}d & -0.0347222 \tabularnewline
\hline 
7  & CSL bilayer  & Fig.~\ref{fig:k5-10}e, \ref{fig:k5-10}f & -0.0273888\tabularnewline
\hline 
8  & 8-site cluster  & Fig.~\ref{fig:k5-10}g, \ref{fig:k5-10}h & -0.0234375 \tabularnewline
\hline 
9  & Inhomogeneous bilayer  & Fig.~\ref{fig:k5-10}i, \ref{fig:k5-10}j & -0.0188265 \tabularnewline
\hline 
10  & CSL bilayer  & Fig.~\ref{fig:k5-10}e, \ref{fig:k5-10}f & -0.01577\tabularnewline
\hline 
\end{tabular}
\caption{Ground state saddle-point patterns of $\chi_{r r'}$, and the corresponding energies
in units of $N{\cal J}N_{s}$ for $k=2,3,\dots,10$. The different types of large-$N$ ground states are described in the text, and depicted in figures as indicated.}
\label{table:summary} 
\end{table}

To fully understand the different ground states, one has to go beyond
the $N = \infty$ or mean-field description. At the mean-field level,
the number of ground state arrangements of clusters or bilayers on the cubic lattice
diverges with the system size.  For example, there are usually many ways to tile the lattice
with a given type of $n$-site cluster.  Also, in the CSL bilayer state, the direction of flux can be chosen independently in each bilayer without affecting the $N = \infty$ ground state energy.
Such degeneracies can be resolved by computing the first correction (perturbative in $1/N$) to the ground state energy;\cite{read1989somefeatures} these calculations are left for future work.

In cluster states, another important effect of fluctuations
is to confine the $f_{r\alpha}$ fermions; the cluster states are thus ``ordinary'' broken symmetry states, without exotic excitations.  A more extensive discussion of fluctuations appears in Ref.~\onlinecite{hermele2011topological}, and the resulting physical properties of the CSL bilayer are discussed in Sec.~\ref{sec:discussion}.  We have not considered the effect of fluctuations in the $k=9$ inhomogeneous bilayer ground state.

\subsection{Detailed descriptions of the mean-field ground states}
\begin{figure}
\includegraphics[scale=0.5]{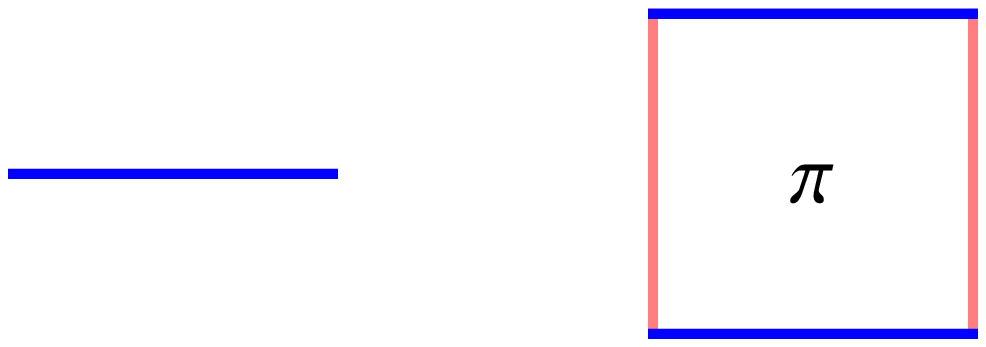}

(a) $k=2$

\medskip{}

\includegraphics[scale=0.5]{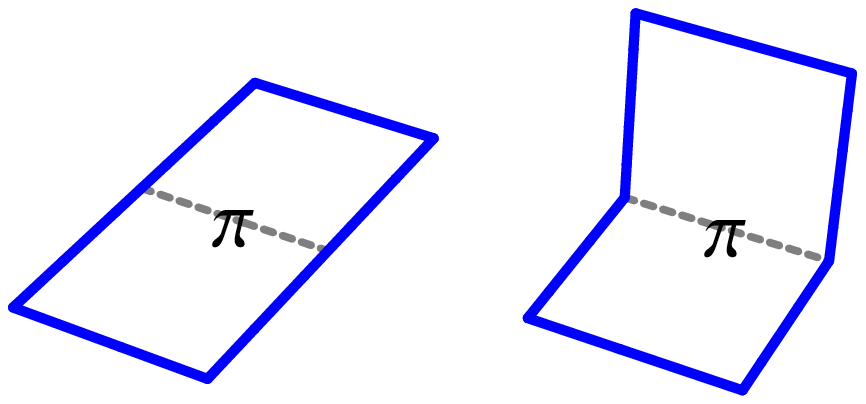}

(b) $k=3$

\medskip{}

\includegraphics[scale=0.5]{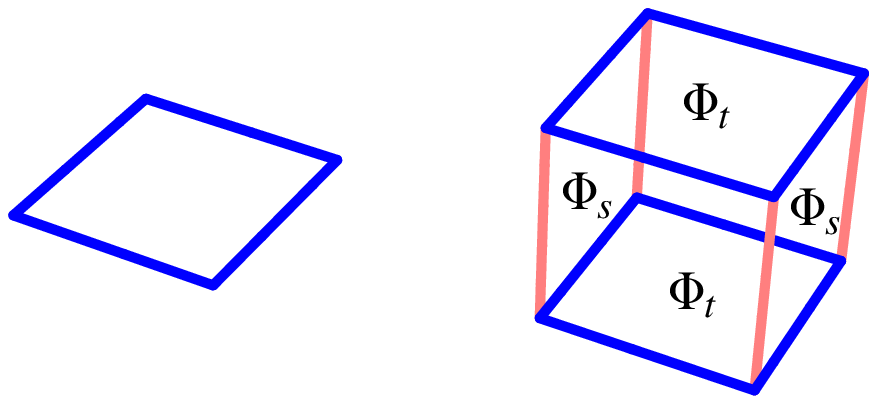}

(c) $k=4$

\caption{Ground-state clusters for $k=2,3,4$.  Shaded bonds are those with
$\chi_{r r'} \neq 0$. Bonds with different shading (or color in online version)
may have different magnitudes $\left|\chi_{\mathbf{rr}'}\right|$.  (a)  The $k=2$ ground state clusters 
are dimers and square plaquettes.  The square plaquette is pierced by $\pi$-flux, and the ratio of $\left|\chi_{\mathbf{rr}'}\right|$ on light (pink online) and dark (blue online) bonds can be chosen arbitrarily.  Setting $|\chi_{r r'}| = 0$ on the two light (pink) bonds breaks the plaquette into two dimers.
(b) The $k=3$ ground state cluster is a 6-site chain pierced by $\pi$-flux.  On the cubic lattice, such chains can exist either as a flat rectangular loop (left), or as the same loop bent by $90^\circ$ in the middle (right). In both cases, $\chi_{r r'} = 0$ on the dashed bond passing through the middle of the loop.  (c) The $k=4$ ground state clusters are square plaquettes and 8-site cubes with $\Phi_{s}$-flux through
the side plaquettes and $\Phi_{t}$-flux through  top and bottom plaquettes. There is a continuous one-parameter family of ground states on an 8-site cube, described in the text.}
\label{fig:cluster_k234}
\end{figure}

We now discuss the mean-field ground states for each value of $k$.  We note that, for $k \geq 5$, we cannot rule out the possibility that the true ground state is lower in energy than the ground state we found.
The ground-state clusters for $k = 2,3,4$ are depicted in Fig.~\ref{fig:cluster_k234}.  These are essentially the same as found in the two-dimensional square lattice,\cite{hermele2009mottinsulators,hermele2011topological} but going to the three-dimensional cubic lattice permits a greater variety of clusters for $k = 3,4$.

It was noted in Ref.~\onlinecite{read1989somefeatures} that for $k=2$ there is actually a \emph{continuous} family of $N = \infty$ ground states, which can be seen for a single square plaquette as shown in Fig.~\ref{fig:cluster_k234}a and discussed in the figure caption.  This continuous ground state degeneracy is also resolved by the order-$1/N$ corrections to the ground state energy.\cite{read1989somefeatures}  We found that a similar continuous degeneracy occurs for $k = 4$ on a single cube (see Fig.~\ref{fig:cluster_k234}c).  As in the figure, consider a single cube with flux $\Phi_t$ through the top and bottom plaquettes (\emph{i.e.}, those lying in the $xy$-plane), and flux $\Phi_s$ through the side plaquettes (\emph{i.e.}, those lying in the $xz$- and $yz$-planes).  Flux passing from the center of the cube to the outside is taken positive.  In order to reach the ground state we must have $2 \Phi_t + 4 \Phi_s = \pm 2\pi$; we choose the positive sign without loss of generality.  We let $\Phi_t = 4 u$ and $\Phi_s = \pi/2 - 2 u$; a ground state is obtained if we restrict $0 \leq u \leq \pi/2$.
In this situation the magnitude $|\chi_{r r'}|$ will generally differ on vertical bonds and other bonds [shaded light (pink) and dark (blue), respectively, in Fig.~\ref{fig:cluster_k234}].  The energy is minimized and saturates the lower bound when
\begin{equation}
\frac{ | \chi_{{\rm light}} | }{ | \chi_{{\rm dark}} | } = 2 \sqrt{ \cos u \sin u } \text{.}
\end{equation}

\begin{figure}
\begin{minipage}[t]{0.45\columnwidth}%
\includegraphics[width=1\columnwidth, height=0.9\columnwidth]{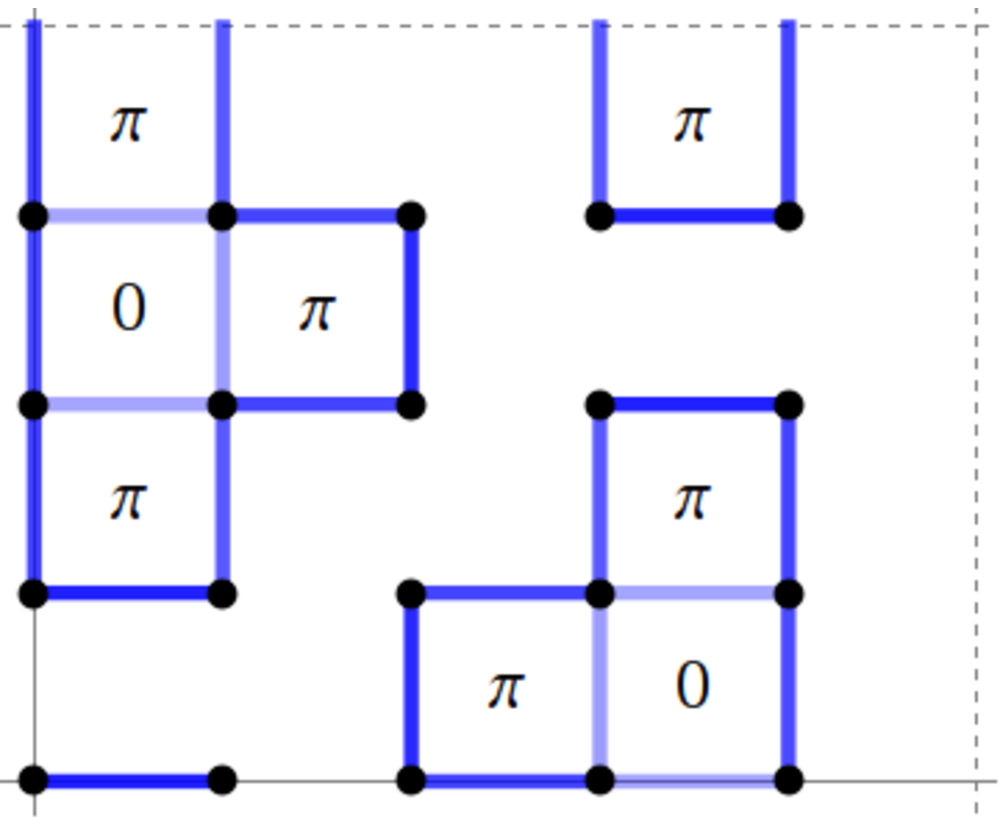}

(a) $k=5$%
\end{minipage} %
\begin{minipage}[t]{0.45\columnwidth}%
\includegraphics[width=1\columnwidth, height=0.9\columnwidth]{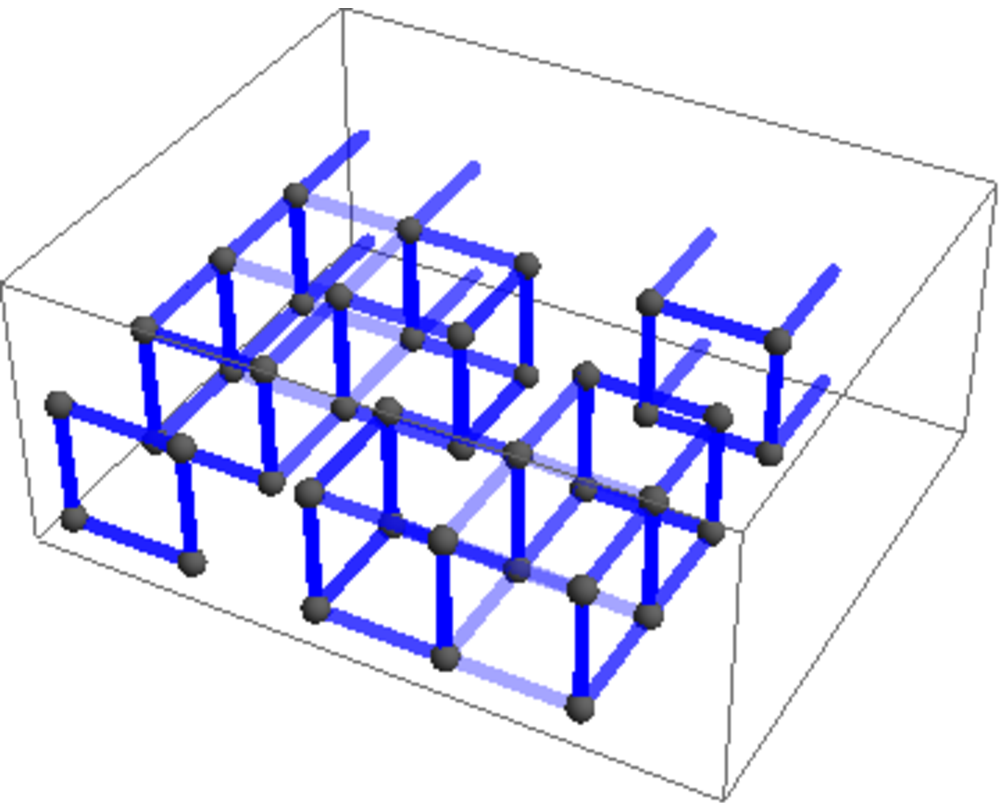}

(b) $k=5$%
\end{minipage}

\medskip{}

\begin{minipage}[t]{0.45\columnwidth}%
\includegraphics[width=1\columnwidth, height=0.7\columnwidth]{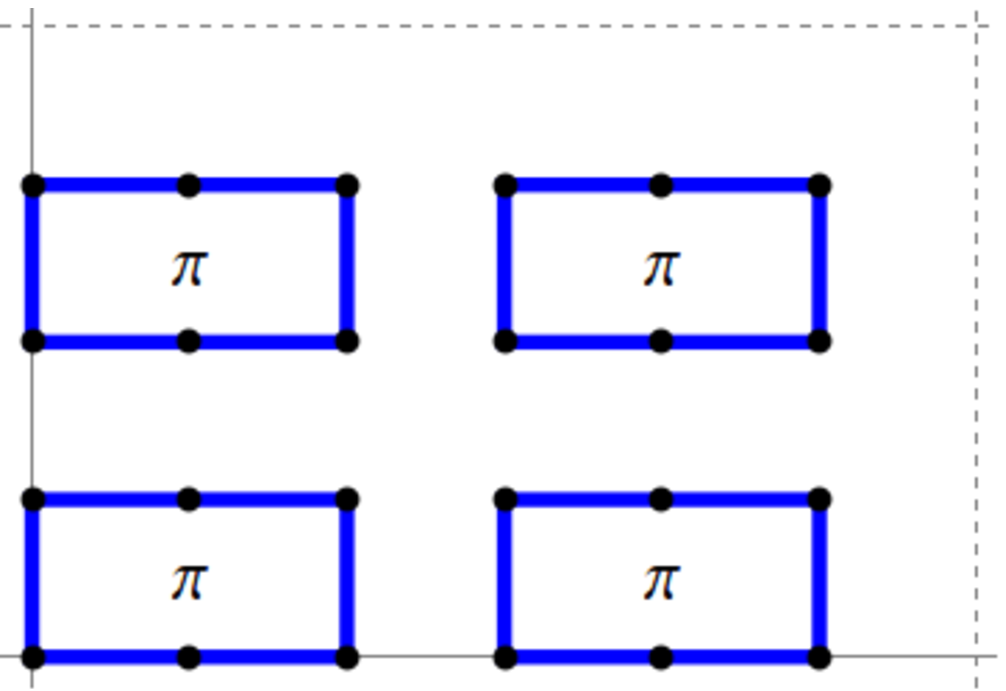}

(c) $k=6$%
\end{minipage} %
\begin{minipage}[t]{0.45\columnwidth}%
\includegraphics[width=1\columnwidth, height=0.7\columnwidth]{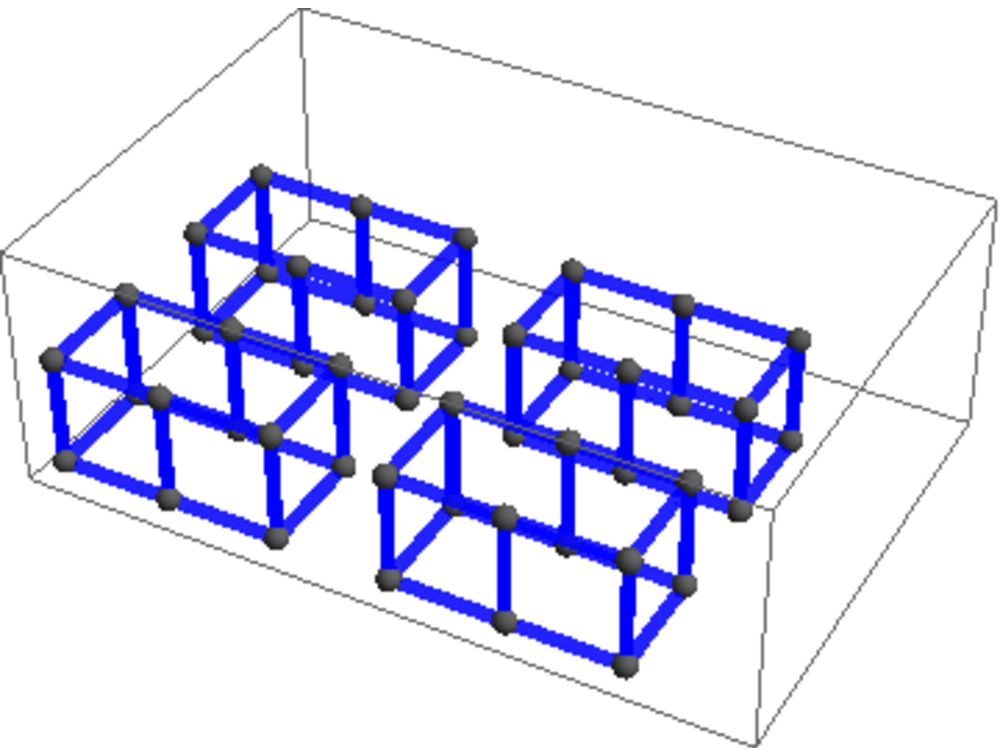}

(d) $k=6$%
\end{minipage}

\medskip{}

\begin{minipage}[t]{0.45\columnwidth}%
\includegraphics[width=1\columnwidth, height=0.75\columnwidth]{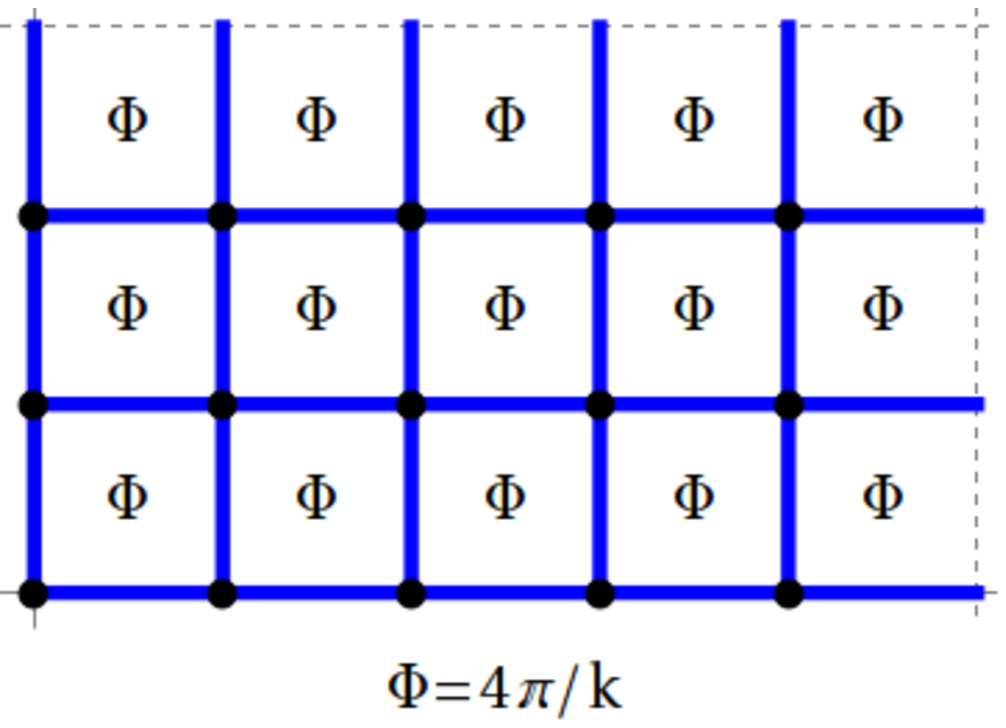}

(e) $k=7,10$%
\end{minipage} %
\begin{minipage}[t][1\totalheight][s]{0.45\columnwidth}%
\includegraphics[width=1\columnwidth, height=0.75\columnwidth]{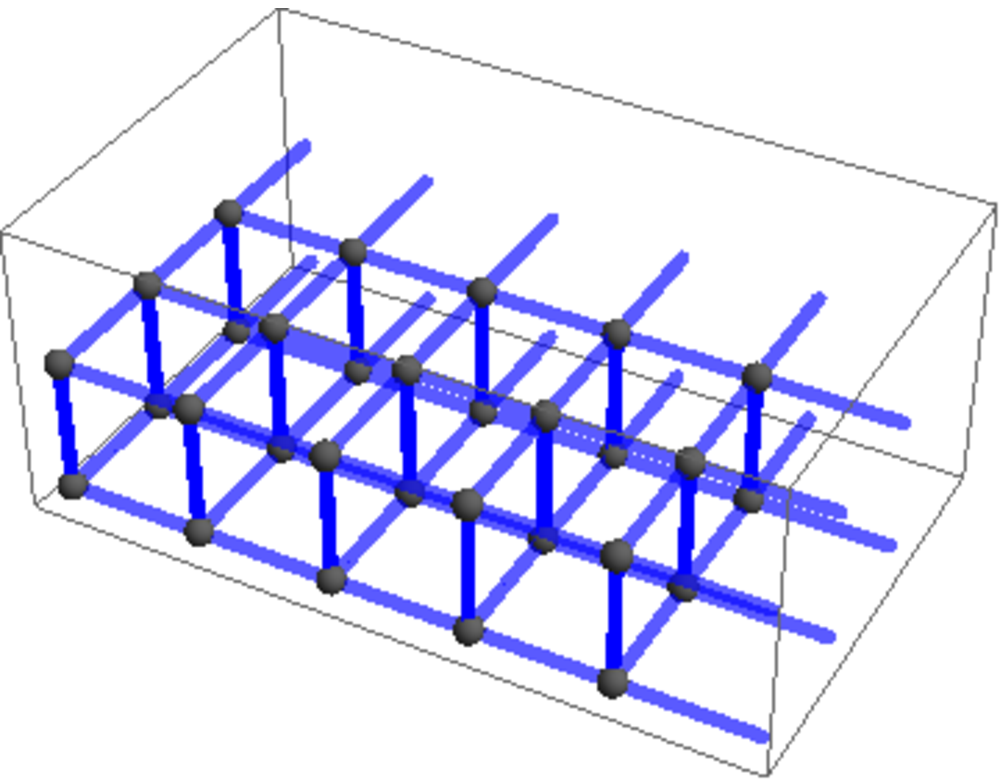}

(f) $k=7,10$%
\end{minipage}

\medskip{}

\begin{minipage}[t][0.2\columnwidth]{0.45\columnwidth}%
\includegraphics[width=0.9\columnwidth, height=0.9\columnwidth]{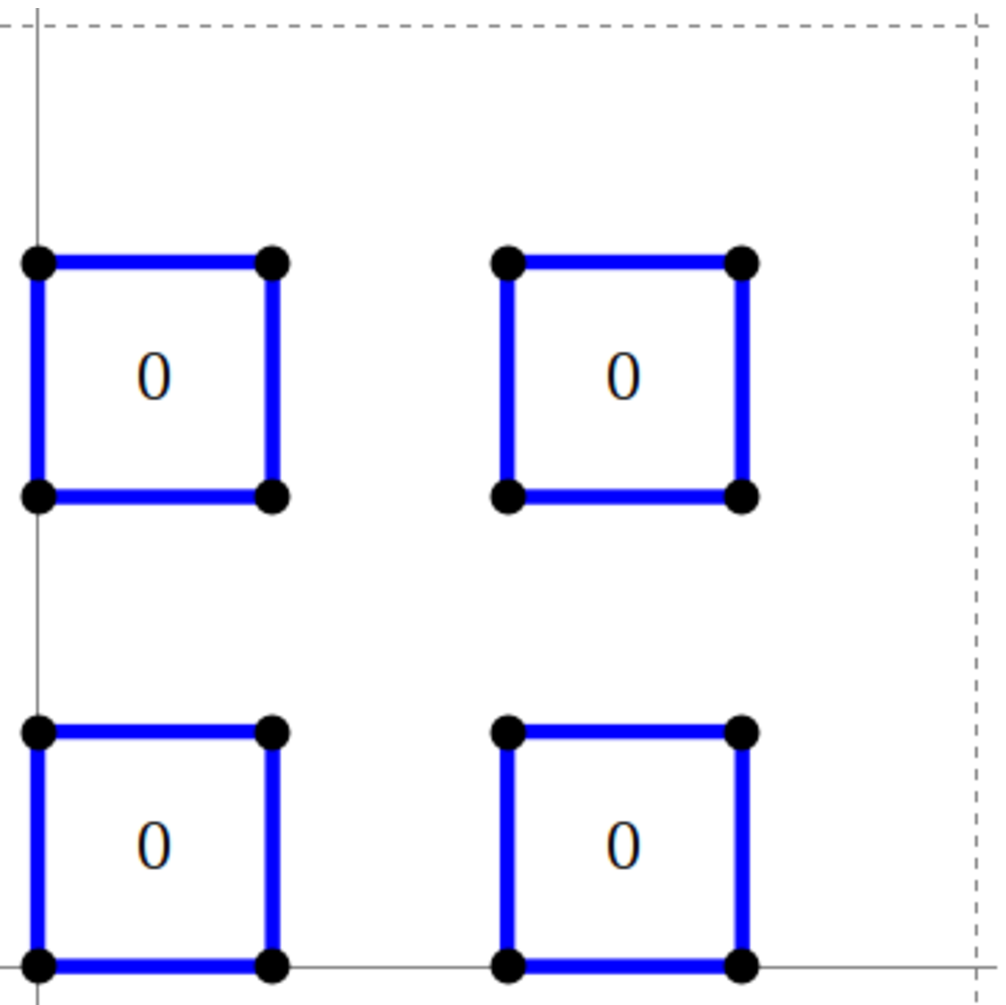}

(g) $k=8$%
\end{minipage} %
\begin{minipage}[t]{0.45\columnwidth}%
\includegraphics[width=1\columnwidth, height=0.9\columnwidth]{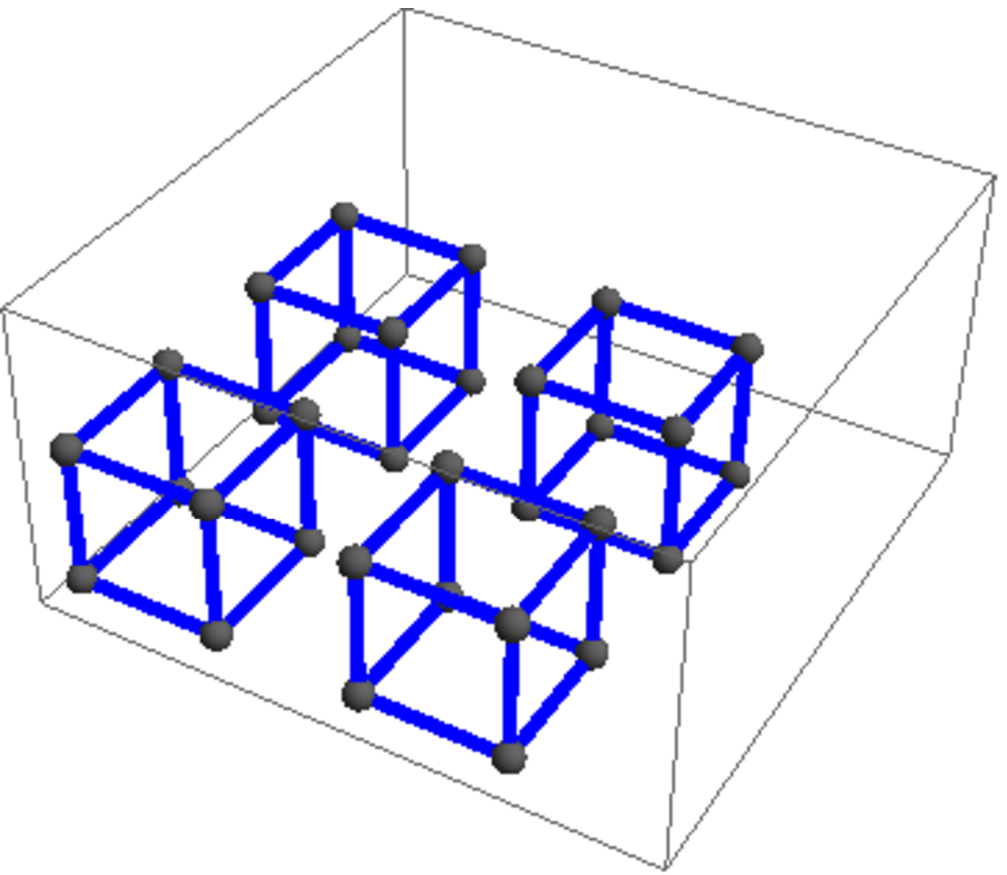}

(h) $k=8$%
\end{minipage}

\medskip{}

\begin{minipage}[t][0.2\columnwidth]{0.45\columnwidth}%
\includegraphics[width=0.9\columnwidth]{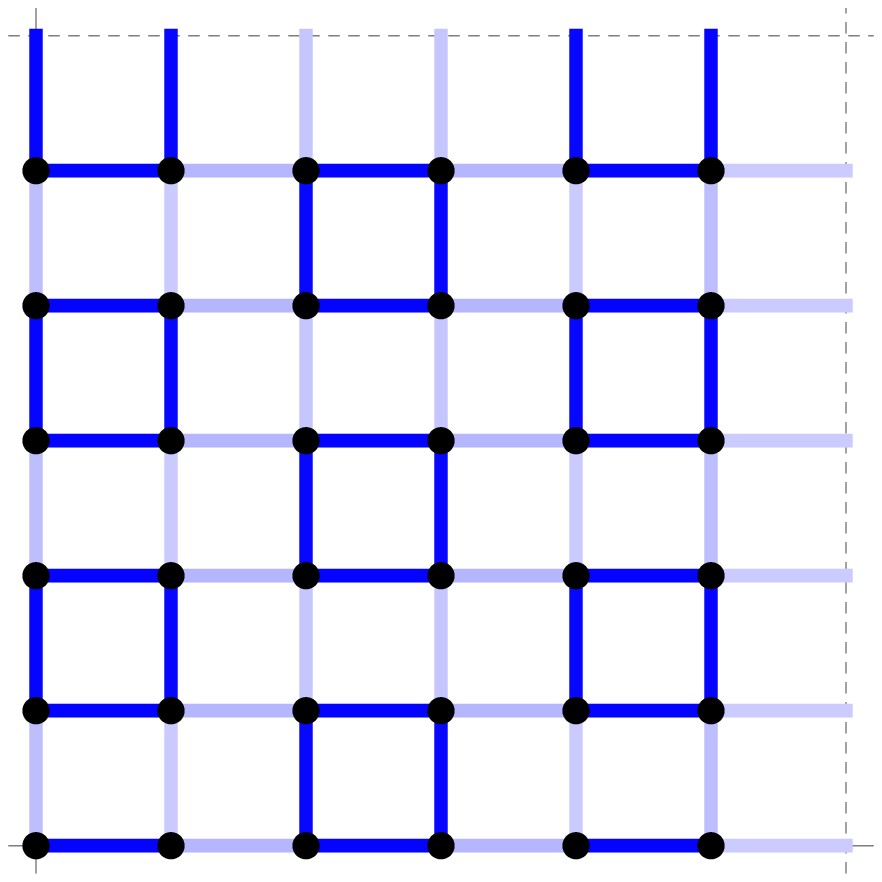}

(i) $k=9$%
\end{minipage} %
\begin{minipage}[t]{0.45\columnwidth}%
\includegraphics[width=1\columnwidth]{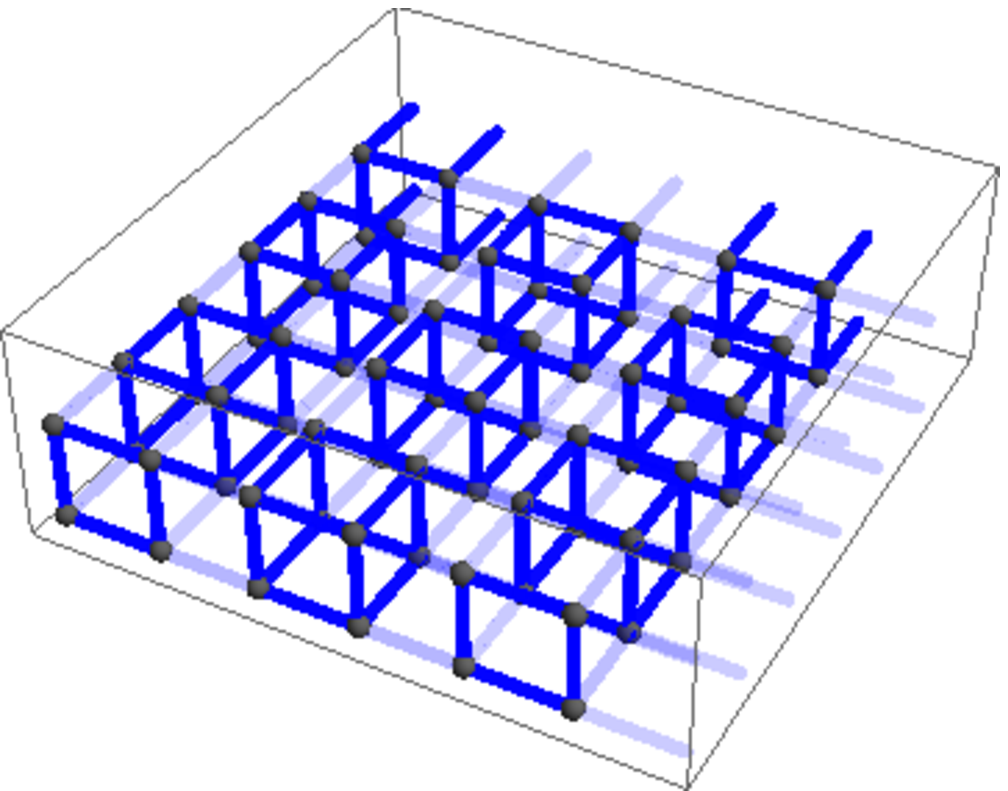}

(j) $k=9$%
\end{minipage}

\caption{Ground-state saddle point configurations of $\chi_{r r'}$ for $k=5,6,\dots,10$. The
right column is a three-dimensional view of each configuration, with larger magnitude
$\left|\chi_{r r'} \right|$ indicated by darker shading.  All these saddle points can be viewed as bilayer structures, with $\chi_{r r'}$ identical on top and bottom layers.  The left column thus shows $|\chi_{r r'}|$ on a single layer, with fluxes indicated except for $k = 9$, where the fluxes are generally non-zero but follow a complicated pattern.  Also, for $k = 5,6$ the fluxes and $|\chi_{r r'}|$ can be changed continuously within a single cluster without affecting the energy; only the simplest configurations are shown.}
\label{fig:k5-10}
\end{figure}

The ground-state patterns of $\chi_{rr'}$ for $5 \leq k \leq 10$ are shown in Fig.~\ref{fig:k5-10}.  For $k = 5,6,8$ we again find cluster ground states.  The case $k=8$ is particularly simple; there, each cluster is a fully symmetric cube with $| \chi_{r r'} |$ constant on every bond, and no flux through the cube faces.  The $k = 5$ and $k = 6$ clusters are conveniently thought of as obtained by stacking two single-layer clusters vertically, and connecting them via the vertical bonds.  For $k = 5$ each cluster is a stack of two ten-site T-shaped objects.  The $k = 6$ clusters are obtained by stacking two $k = 3$ ground state clusters (see Fig.~\ref{fig:cluster_k234}b).  In the $k = 5, 6$ cases, our numerical calculations find evidence for a continuous family of degenerate ground states within each cluster, as for the 4-site $k=2$ clusters and 8-site $k=4$ clusters (Fig.~\ref{fig:cluster_k234}).  Unlike in those cases, however, we have not been able to find a simple parametrization of the degenerate ground states.

For $k = 7,9,10$, we find bilayer ground states, with the CSL bilayer saddle point described above occurring for $k = 7, 10$.  The $k = 9$ ground state is more complicated, spontaneously breaking translation symmetry within each bilayer. Time reversal symmetry is broken as well by a complicated pattern of fluxes.  It is interesting to note that \emph{all} the $5 \leq k \leq 10$ ground states have a bilayer structure, as the clusters for $k = 5,6,8$ can be arranged into bilayers (see right column of Fig.~\ref{fig:k5-10}).  In addition, the two square lattice layers of each bilayer have identical $\chi_{r r'}$,  there is zero flux on the ``vertical'' plaquettes connecting the two layers, and the vertical bonds have magnitude $|\chi_{r r'} | = {\cal J} / k$. \cite{tuning} As discussed below, this simple structure allows us to exploit a useful relation with the single-layer square lattice at filling parameter $k' = k/2$.

\subsection{Obtaining the mean-field results}
We now describe how the large-$N$ ground states were determined. As on the square lattice,\cite{hermele2009mottinsulators,hermele2011topological} the results for $k=2,3,4$ are rigorous, and are obtained by applying a lower bound on $E_{MFT}$ obtained by Rokhsar for $k=2$,\cite{rokhsar1990quadratic} and generalized to $k > 2$ (with a stronger bound holding for bipartite lattices) in Refs.~\onlinecite{hermele2009mottinsulators,hermele2011topological}.  Cluster states for $k=2,3,4$ on the square\cite{hermele2009mottinsulators,hermele2011topological}  and cubic lattices saturate this lower bound.  A necessary condition for saturation on a bipartite lattice is that the mean-field single-particle energy spectrum must be completely flat, with only three energies $0, \pm \epsilon$ occuring in the spectrum, and with energy $-\epsilon$ states filled and others empty.\cite{hermele2009mottinsulators,hermele2011topological}  We believe that this kind of spectrum can only be produced by a cluster state.  Moreover, for larger clusters (and thus with increasing $k$), it becomes harder to arrange for a spectrum containing only three energies.  While we do not have a rigorous proof, we believe saturation is impossible for $k > 4$ on the square and cubic lattices.

For $k\geq5$, we resort to a numerical approach to find the ground states.  We employ
the self-consistent minimization (SCM) algorithm developed in Refs.~\onlinecite{hermele2009mottinsulators,hermele2011topological}, which proceeds as follows (see Ref.~\onlinecite{hermele2011topological} for more details):
\begin{algorithmic}[1]
\STATE Start with $\mu_{r}=0$ and a randomly generated configuration of
$\chi_{rr'}$. 
\STATE Adjust $\mu_r$ to satisfy the saddle-point equation
\begin{equation}
\langle f_{r}^{\alpha\dagger}f_{r\alpha} \rangle =m,  \quad \text{for all } r\label{eq:fix_mu} \text{.}
\end{equation}
$\mu_{r}$ is determined by a multidimensional Newton's method.\cite{hermele2009mottinsulators,hermele2011topological,numericalrecipes}  Stop if no solution is found.
\STATE Generate a new $\chi_{r r'}$  using the saddle-point equation
\begin{equation}
\chi_{rr'}=-\frac{{\cal J}}{N}\left\langle  f_{r'}^{\alpha\dagger}f_{r\alpha} \right\rangle .\label{eq:iteration_chi}
\end{equation}
\STATE Go back to step 2 until $\chi_{r r'}$ and $\mu_r$ converge.
\end{algorithmic}
As long as step 2 is successful, the energy $E_{MFT}$ is guaranteed to decrease with each iteration of the SCM algorithm.\cite{hermele2009mottinsulators,hermele2011topological}   But a random initial configuration
of $\chi_{r r'}$ does not necessarily converge to the ground state, and can
instead converge to a local minimum of $E_{MFT}$.
Therefore, in order to find the ground state, we need
to try as many independent random initial configurations of $\chi_{r r'}$ as possible.  For those random initial configurations resulting in the lowest energies, we found extremely good convergence in $E_{MFT}$ by the time the SCM procedure is stopped (typically after 300 iterations), and effects of  randomness on the reported values of $E_{MFT}$ are thus entirely negligible.

\begin{figure}
\begin{minipage}[t]{0.45\columnwidth}%
\includegraphics[width=1\columnwidth]{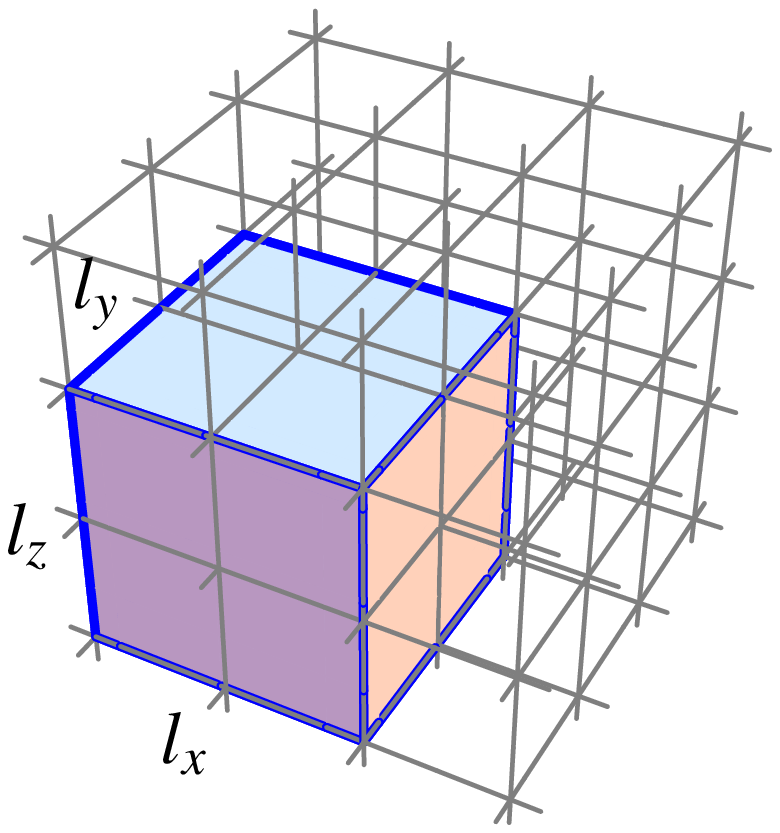}

(a)%
\end{minipage}%
\begin{minipage}[t]{0.45\columnwidth}%
\includegraphics[width=1\columnwidth]{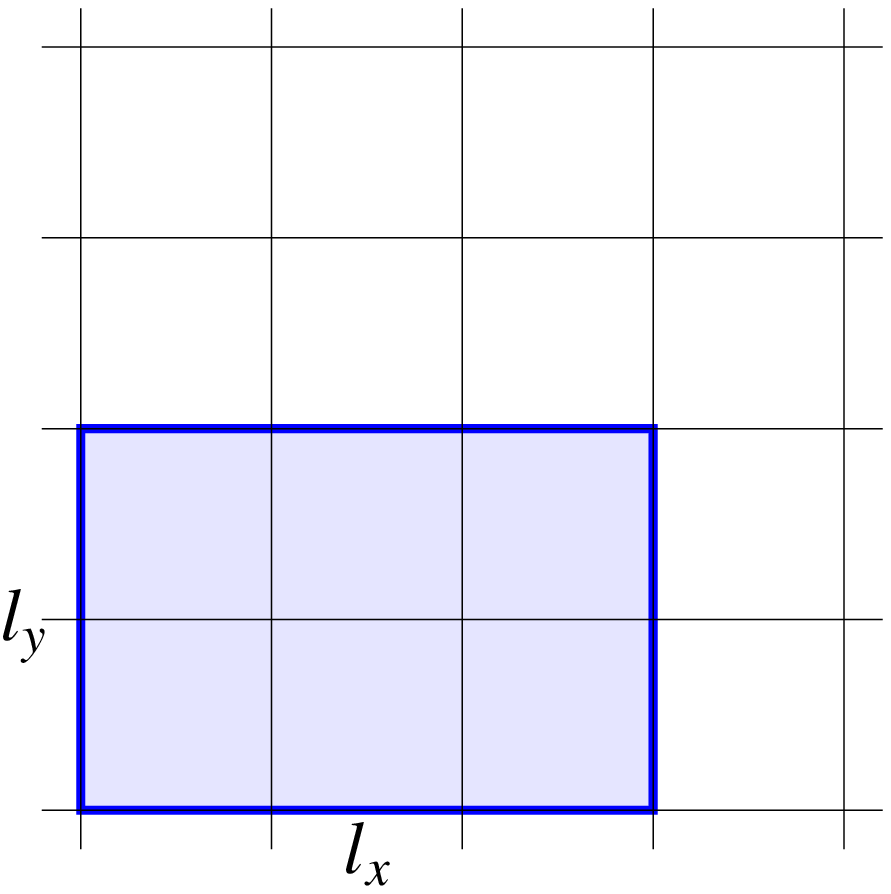}

(c)%
\end{minipage}

\begin{minipage}[t]{1\columnwidth}%
\includegraphics[width=0.7\columnwidth]{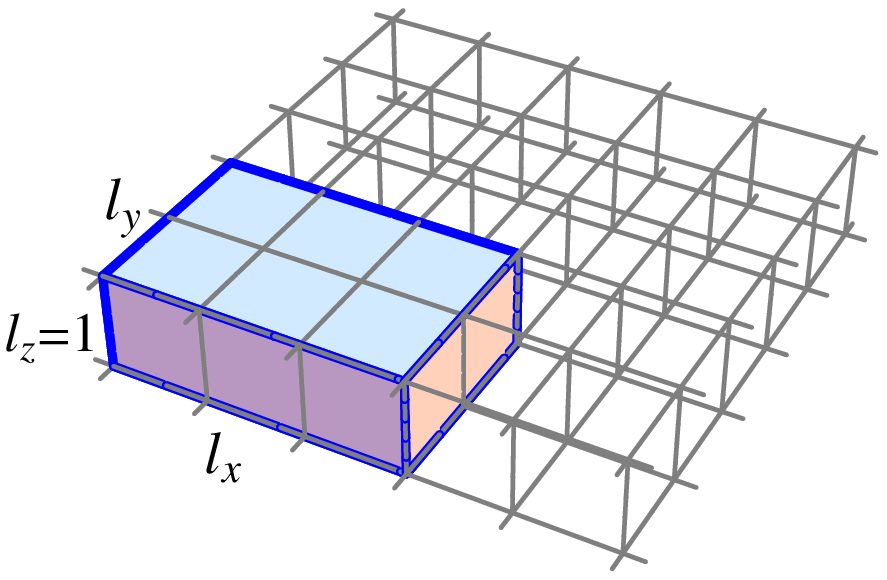}

(b)%
\end{minipage}

\caption{Unit cells used for SCM calculations on the cubic lattice (a), single bilayer (b), and single-layer square lattice (c).  In the cubic case the primitive Bravais lattice vectors are chosen parallel to the edges of the rectangular prismatic unit cell.  The analogous statement is true for the bilayer and single-layer cases, with primitive Bravais lattice vectors parallel to the $l_{x,y}$ edges of the unit cell.}
\label{fig:cells}

\end{figure}

To improve the performance of the SCM algorithm, we define $\chi_{r r'}$
with $\mu_r$ within some fixed unit cell, which is then repeated periodically
to cover a finite-size  $L_x \times L_y \times L_z$ lattice with periodic boundary conditions.  For simplicity, we always
choose the unit cell to be a rectangular prism with edge lengths $l_{x,y,z}$ (see Fig.~\ref{fig:cells}),  with primitive Bravais lattice vectors parallel to the edges of the rectangular prism.\cite{prismnotes}   For each value of $k$, we choose the minimum linear system size $L = \operatorname{min}(L_x, L_y, L_z)$ to be as large as possible given the constraints of our available computing resources and the need to try a reasonably large number of different random initial conditions.  In some cases we also considered larger system sizes, especially when we found competing saddle points very close in energy.  A more careful study of finite-size effects would be desirable, but due to the above constraints we leave this for future work.  Table~\ref{tab:Unit-cells} displays the range of unit cell dimensions studied for each value of $k$, as well as the number of random initial conditions tried for each cell, and the minimum linear system size $L$.

\subsection{Relation between bilayer states and square lattice saddle points}
As noted above, the ground states for $5 \leq k \leq 10$ can all be viewed as bilayer states, which means that such saddle points can also be obtained by a studying the large-$N$ Heisenberg model on a single bilayer.  We have also carried out SCM numerical calculations in this geometry (see Table~\ref{tab:Unit-cells} and Fig.~\ref{fig:cells} for more information); this is computationally cheaper than the cubic lattice SCM calculations, and provides a useful check on those results.  These bilayer SCM calculations find the same ground states as the corresponding cubic lattice calculations, except for $k = 9$, where the bilayer calculation finds a lower-energy state that can then be extended to a cubic lattice saddle point.  Presumably, this saddle point would also be found by SCM on the cubic lattice with enough runs using independent random initial conditions.

There is an interesting relation between certain saddle points of a single bilayer, and corresponding saddle points of a single-layer square lattice, but with filling parameter $k' = k/2$.  The cubic lattice ground states for $5 \leq k \leq 10$ are all of this type.  We label the sites of a single bilayer by $(r, i)$, where $i = 1,2$ is the layer index, and $r$ labels the square lattice sites within each layer.  There are $N_s = 2 N^{2d}_s$ lattice sites, where $N^{2d}_s$ is the number of sites in a single layer.  Consider a saddle point where
\begin{eqnarray}
\chi_{r1, r'1} = \chi_{r2, r'2} &\equiv& \chi_{r r'} \\
\mu_{r1} = \mu_{r2} &\equiv& \mu_r \\
\chi_{r1, r2} \equiv \chi_v \text{.}
\end{eqnarray}
Here, $\chi_v$ is real and positive, and all other inter-layer $\chi$'s are assumed to vanish.  We let $n$ label the one-particle eigenstates of a single layer, with energies $\epsilon^{2d}_n$.  The full one-particle spectrum is then given by
\begin{equation}
\epsilon_{n, \sigma} = \epsilon^{2d}_n + \sigma \chi_v \text{,}
\end{equation}
where $\sigma = \pm 1$.  We assume that the energy spectrum and filling are such that only $\sigma = -1$ states are occupied by fermions, in which case the two-dimensional spectrum $\epsilon^{2d}_n$ (shifted in energy by $-\chi_v$) is filled by $N N_s / k = 2 N N^{2d}_s / k$ fermions.  This corresponds to a single-layer problem  with twice as many fermions, or filling parameter $k' = k/2$.  The saddle point energy is then
\begin{eqnarray}
E_{MFT} &=& N N^{2d}_s \frac{\chi^2_v}{{\cal J}} - \frac{2 N N^{2d}_s}{k} \chi_v  \label{eqn:bilayer-emft}\\
&+& \nonumber \frac{2 N}{{\cal J}} \sum_{\langle r r' \rangle} | \chi_{r r'} |^2 + m' \sum_r \mu_r + E^{2d}_f(k') \text{.}
\end{eqnarray}
Here, $m' = 2 m$, and $E^{2d}_f(k')$ is the ground state energy of the fermionic part of the mean-field Hamiltonian [last two terms of Eq.~(\ref{eq:MFT})], for a single-layer square lattice with filling parameter $k'$.  The first two terms of Eq.~(\ref{eqn:bilayer-emft}) are minimized with respect to $\chi_v$ to find $\chi_v =  {\cal J} / k$.  The last three terms combine to $E^{2d}_{MFT}(k', {\cal J}')$, the saddle point energy of a single-layer square lattice with filling parameter $k'$ and ${\cal J}' = {\cal J}/2$.  Noting that
\begin{equation}
E^{2d}_{MFT}(k') \equiv E^{2d}_{MFT}(k', {\cal J}) = 2 E^{2d}_{MFT}(k', {\cal J}') \text{,}
\end{equation}
we obtain the following relation between bilayer and single-layer saddle point energies:
\begin{equation}
\frac{E_{MFT}}{N_s N} = - \frac{{\cal J}}{2 k^2} + \frac{1}{4} \frac{E^{2d}_{MFT}(k/2) }{N^{2d}_s N} \text{.}
\end{equation}

This relation allows us to study via SCM the single-layer square lattice with filling parameter $k' = k/2$ as a further check on the cubic lattice results.  For integer $k'$, this was already done in Ref.~\onlinecite{hermele2009mottinsulators}.  We carried out SCM calculations for the half-odd integer filling parameters $k'=\frac{5}{2},\frac{7}{2},\frac{9}{2}$ (see Table~\ref{tab:Unit-cells} and Fig.~\ref{fig:cells}).  For all values of $k$, 
these calculations find the same ground states as found by the single-bilayer SCM calculations.  

As a further check on our results, we also computed the energies of some simple competing states.  Table~\ref{tab:Mean-field-energies} compares the energies of these states to the ground state saddle point energies found by SCM.

\begin{table}
\begin{tabular}{|c|c|c|c|c|c|c|}
\hline 
$k$ & \multicolumn{2}{c|}{Cubic lattice} & \multicolumn{2}{c|}{Single bilayer} & \multicolumn{2}{c|}{$k/2$ square lattice}\tabularnewline
\hline 
\hline 
5 & $1\leq l_{x,y,z}\leq5$ & 10 & $1\leq l_{x,y}\leq5$ & 10 & $1\leq l_{x,y}\leq6$ & 30\tabularnewline
 &  & \emph{30} &  & \emph{60} &  & \emph{60}\tabularnewline
\hline 
6 & $1\leq l_{x,y,z}\leq6$ & 4 & $1\leq l_{x,y}\leq6$ & 4 &  & \tabularnewline
 &  & \emph{30} &  & \emph{60} &  & \tabularnewline
\hline 
7 & $1\leq l_{x,y,z}\leq7$ & 4 & $1\leq l_{x,y}\leq7$ & 10 & $1\leq l_{x}\leq7$ & 20\tabularnewline
 &  & \emph{21} &  & \emph{35} & $1\leq l_{y}\leq10$ & \emph{42}\tabularnewline
\hline 
8 & $1\leq l_{x,y}\leq8$ & 4 & $1\leq l_{x,y}\leq8$ & 4 &  & \tabularnewline
 & $1\leq l_{z}\leq5$ & \emph{24} &  & \emph{40} &  & \tabularnewline
\hline 
9 & $1\leq l_{x,y}\leq9$ & 4 & $1\leq l_{x}\leq9$ & 10 & $1\leq l_{x}\leq10$ & 10\tabularnewline
 & $1\leq l_{z}\leq4$ & \emph{36} & $1\leq l_{y}\leq11$ & \emph{36} & $1\leq l_{y}\leq9$ & \emph{36}\tabularnewline
\hline 
10 & $1\leq l_{x,y}\leq10$ & 4 & $1\leq l_{x,y}\leq10$ & 5 &  & \tabularnewline
 & $1\leq l_{z}\leq4$ & \emph{30} &  & \emph{60} &  & \tabularnewline
\hline 
\end{tabular}

\caption{This table contains information about our SCM numerical study on the cubic lattice (1st column), as well as the related problems of a single bilayer (2nd column), and single layer square lattice with $k' = k/2$ (3rd column).  On the left-hand side of each entry of the table, the range of unit cell dimensions is shown as an inequality.  For every choice of $l_{x,y,z}$ within the given range, the number of times we ran the SCM algorithm with distinct random initial configurations of $\chi_{r r'}$ is shown on the right-hand side of the entry (top).  Also on the right-hand side is the minimum linear system size $L$ (bottom, italics).  \label{tab:Unit-cells}}
\end{table}

\begin{table*}

\begin{tabular}{|c|c|c|c|c|c|c|}
\hline 
$k$ & 5 & 6 & 7 & 8 & 9 & 10\tabularnewline
\hline 
\hline 
Bilayer ($\Phi = 2\pi n / k$) & \textcolor{black}{-0.0444916} & \textcolor{black}{-0.0344012} & \textbf{\textcolor{black}{-0.0273888}} & \textcolor{black}{-0.0223613} & \textcolor{black}{-0.0186271} & \textbf{\textcolor{black}{-0.01577 }}\tabularnewline
\hline 
$k$-site cluster & \textcolor{black}{-0.04} & \textcolor{black}{-0.032407} & \textcolor{black}{-0.026239} & \textbf{\textcolor{black}{-0.0234375 }} & \textcolor{black}{-0.0178326} & \textcolor{black}{-0.014 }\tabularnewline
\hline 
Uniform real $\chi$ & \textcolor{black}{-0.0394159} & \textcolor{black}{-0.0312776} & \textcolor{black}{-0.0254048 } & \textcolor{black}{-0.0210391 } & \textcolor{black}{-0.0177088} & \textcolor{black}{-0.0151133}\tabularnewline
\hline 
$\left(2\pi n_{x,y,z}/k\right)$-flux & \textcolor{black}{-0.0430802} & \textcolor{black}{-0.0330693 } & \textcolor{black}{-0.0261299 } & \textcolor{black}{-0.0212772 } & \textcolor{black}{-0.0177579 } & \textcolor{black}{-0.0151134 }\tabularnewline
\hline 
SCM ground state & \textbf{\textcolor{black}{-0.0445021}} & \textbf{\textcolor{black}{-0.0347222 }} & \textbf{\textcolor{black}{-0.0273888}} & \textbf{\textcolor{black}{-0.0234375 }} & \textbf{\textcolor{black}{-0.0188265}} & \textbf{\textcolor{black}{-0.01577}}\tabularnewline
\hline 
\end{tabular}
\caption{Comparison of energies of a variety of simple saddle points (top four rows), with the energy of the ground state found by SCM numerics (bottom row).  All energies are in units of $N {\cal J} N_s$.  Each row represents a class of saddle points, described below.  For classes including multiple different saddle points, the energy shown is the lowest in the class.  We considered the following classes of saddle points: \emph{Bilayer} ($\Phi = 2\pi n / k$). We considered a generalization of the CSL bilayer saddle point described in the main text, where the flux through each plaquette is $\Phi = 2 \pi n / k$, where $n = 0, \dots, k-1$.  \emph{$k$-site cluster}.  The energy of a cluster with $k$ sites is proportional to the number of bonds in the cluster,\cite{hermele2009mottinsulators,hermele2011topological} so the lowest-energy such state can be found by finding a $k$-site cluster containing the greatest number of bonds.  \emph{Uniform real $\chi$}.  This is the state where $\chi_{r r'}$ is real and spatially constant.  \emph{$(2\pi n_{x,y,z} / k )$-flux}.  These states have $2 \pi n_x / k$ flux through every plaquette normal to the $x$-direction, and similarly for $y$ and $z$, where $0 \leq n_{x,y,z} \leq k-1$.  Since most of these states break lattice rotation symmetry, the magnitude $|\chi_{r r'}|$ is allowed to vary depending on bond orientation, but is fixed to be translation invariant.\cite{specialsaddle}}
\label{tab:Mean-field-energies}
\end{table*}

\section{Discussion}
\label{sec:discussion}

The large-$N$ results presented here find a rich variety of candidate non-magnetic ground states for Mott insulators of ultra-cold fermionic AEA.  It would be fascinating to realize any of these states experimentally.  In order to achieve this, there still need to be substantial advances in preparation of low-entropy magnetic states of ultra-cold atoms, and our results add to the increasing motivation to pursue such advances specifically in AEA systems.  In addition, if future experiments can enter a regime where any of the states discussed here can be realized, it will be of crucial importance to devise probes of their characteristic properties.

We would like to close by highlighting the CSL bilayer state, which has some striking properties that would be fascinating to realize experimentally, and which we now briefly discuss.  At the large-$N$ mean-field level the cubic lattice breaks into disconnected bilayers, and one can understand the properties beyond mean-field theory by first focusing on a single bilayer.  The effect of fluctuations is to couple the fermions to a dynamical ${\rm U}(1)$ gauge field.  The mean-field fermions are in a gapped integer quantum Hall state, so integrating them out generates a Chern-Simons term for the ${\rm U}(1)$ gauge field.  Because the mean-field fermions in a single bilayer and in the single-layer square lattice CSL\cite{hermele2009mottinsulators,hermele2011topological}  have in both cases a single chiral edge mode per spin species, the coefficient of the Chern-Simons term and associated topological properties are the same.  The spinons are Abelian anyons with statistics angle $\theta = \pi \pm \pi / N$, and there is a chiral edge mode with gapless excitations carrying $SU(N)$ spin, which is described by a chiral $SU(N)_1$ Wess-Zumino-Witten model.\cite{hermele2009mottinsulators,hermele2011topological}

If adjacent bilayers are coupled weakly, bulk properties are unaffected due to the energy gap.  One simply has a many-layer CSL state, with anyonic spinons confined to the the individual bilayers.  Due to the gapless edge modes of single bilayers, the physics on the two-dimensional surface is likely more interesting.  This depends crucially on whether adjacent bilayers have the same or opposite magnetic flux, as the direction of the flux controls the direction of the chiral edge modes.  If the fluxes are aligned oppositely in neighboring bilayers, then edge modes on neighboring bilayers are counterpropagating and   an energy gap is possible on the two-dimensional surface.  On the other hand, if all fluxes are parallel, then all the chiral edge modes propagate in the same direction, and the two-dimensional surface is expected to remain gapless.  The resulting surface state is a kind of two-dimensional chiral ``spin metal,'' which could be interesting to study in future work.

\acknowledgments

M.H. gratefully acknowledges Victor Gurarie and Ana Maria Rey for related prior collaborations. This work is supported by DOE  award no. DE-SC0003910.

\bibliography{reference}

\begin{thebibliography}{80}
\bibitem [{bos()}]{bosonicnote}%
  \BibitemOpen
  \href@noop {} {}\bibinfo {note} {We do not consider bosonic AEAs, because all
  stable bosonic atoms with similar electronic structure have zero nuclear
  spin.}\BibitemShut {Stop}%
\bibitem [{tun()}]{tuning}%
  \BibitemOpen
  \href@noop {} {}\bibinfo {note} {For $k=5,6$, where we find a continuous
  ground state degeneracy within each cluster, these statements hold for a
  particular choice of ground state in each cluster.}\BibitemShut {Stop}%
\bibitem [{pri()}]{prismnotes}%
  \BibitemOpen
  \href@noop {} {}\bibinfo {note} {We also considered some more general
  parallelepiped unit cells, but less systematically than the rectangular prism
  case.}\BibitemShut {Stop}%
\bibitem [{spe()}]{specialsaddle}%
  \BibitemOpen
  \href@noop {} {}\bibinfo {note} {The calculation procedure is to set up
  special initial data of $\chi_{rr'}$ for iteration under the SCM procedure.
  For those $\chi_{r r'} \neq 0$ in a given state, we choose
  $\left|\chi_{rr'}\right|=0.1$, with the required pattern of fluxes. After
  finding the lowest energy, we checked whether the flux pattern is preserved.
  For $\left(2\pi n_{x,y,z}/k\right)$-flux states, when $k\geq7$ the
  computation time becomes large, so we restricted the range of flux to $0 \leq
  n_{x} \leq 1$, and $0 \leq n_{y,z} \leq k-1$.}\BibitemShut {Stop}%
\end{thebibliography}

\end{document}